# Cross-over mechanism of the melting transition in monolayers of alkanes adsorbed on graphite and the universality of energy scaling.

L. Firlej[1,4], B. Kuchta[1,3], M.W. Roth[2], and C. Wexler[1]

[1]*University of Missouri, Department of Physics and Astronomy, Columbia, MO 65211*

[2]*University of Northern Iowa, Department of Physics, Cedar Falls, IA 50614*

[3]*Laboratoire Chimie Provence, Université Aix-Marseille 1, Marseille, France*

[4]*Laboratoire des Colloides, Verres et Nanomatériaux, Université Montpellier 2, Montpellier, France*

**Abstract**

The interplay between the torsional potential energy and the scaling of the 1-4 van der Waals and Coulomb interactions determines the stiffness of flexible molecules. In molecular simulations often ad-hoc values for the scaling factor (SF) are adopted without adequate justification. In this letter we demonstrate for the first time that the precise value of the SF has direct consequences on the critical properties and mechanisms of systems undergoing a phase transition. By analyzing the melting of n-alkanes (hexane C6, dodecane C12, tetracosane C24) on graphite, we show that the SF is not a universal feature, that it monotonically decreases with the molecular length, and that it drives a cross-over between two distinct mechanisms for melting in such systems.

During the last half of century computational physics has grown in scope and importance to a point where it became a third part of the traditional division between experimental and theoretical physics. Computer simulations brought a remarkable insight into the behavior exhibited by complex systems with large numbers of degrees of freedom. Recent advances in computer power and algorithms have allowed detailed simulations of systems comprised of thousands to millions of atoms, for periods of ns to $\mu$s, permitting detailed pictures of diverse phenomena, from simple atomic processes to complex behavior of biological macromolecules.

The quality of the results of computer simulations of any real system depends, however, on the quality of force fields of the theoretical model. This becomes critically important when modeling large systems of flexible molecules, such as biomolecules (especially proteins or lipids) or polymers, for which alkanes are prototypes. In such systems, correlations between internal and external degrees of freedom determine the local conformational stability of molecules (i.e., folding). Conversely the same correlations affect intermolecular correlated processes (e.g., phase transitions). Being able to correctly account for the energy and ordering of conformations is essential if force field methods are to be considered as *predictive*.

Recently, a number of auto-assemblies for alkanes on graphite have been studied by Molecular Dynamics (MD) methods. [1-7]. Two important conclusions emerged from comparative analyses of those studies. First, an all-atom representation of the molecules was always necessary to reproduce the intricacies between structural and melting properties of the systems [3,6]. Second, it is of extreme importance to model correctly the internal non-bonded interactions because they are the ones that define the molecular stiffness and hence the ability of molecules to deform. The relative stability of a particular conformation of a given molecule is determined by a balance between the torsional energy and the nonbonded energy terms (electrostatic, Van der Waals and repulsive). When the conformational changes are involved in thermodynamic phenomena (such as phase transitions), the necessary precision of the interaction model is even more demanding, as it has to reproduce correctly the whole distribution of instantaneous configurations. This requirement is usually satisfied by a parameterization of the force fields. However, in spite of the enormous progress made in recent years, uncertainties in the non-bonded internal energy components and electrostatic interactions (including the polarization energy) still persist. Fitting force-field parameters is an elaborate task. All of the most popular modern force fields (OPLS-AA [8], CHARMM22 [9], AMBER-94 [10]) define several parameters that attempt to quantify the relative contribution of different internal and intermolecular energy terms to the total energy of the systems. Certainly, it is always desirable for parameters to be transferable and applicable to a wide class of molecules—in practice this is quite difficult to attain.

One way to estimate the universality of adjustable parameters is to check how different components affect the kinetic and thermodynamic quantities that can be extracted from numerical simulations. In most non-polar organic systems, electrostatic terms have a negligible impact [11] and are usually totally ignored [12,13]. However, *ab initio* calculations indicate significant charge separation within a C-H bond (0.3-0.6 D). The OPLS-AA (Optimized Parameters for Liquid Simulations—All Atoms representation) force field tries to account for that, introducing partial charges specific to atoms involved in a given chemical bond. These non-negligible bond dipoles noticeably affect the interaction energies when two molecules approach each other, becoming more important as molecules become less symmetrical.

The modeling of the internal non-bonded interactions is much less obvious: in particular, how to correctly account the 1-4 interactions (atoms separated by three bonds) that are partially included in the dihedral torsion term, i.e., how to avoid *double-counting* those. This is crucial since, as will be shown later in the paper, the interplay between the torsional potential and the scaling of 1-4 van der Waals (vdW) and 1-4 Coulomb interactions determines the internal (conformational) stiffness of the molecule and its ability to deform in a very sensitive way. To deal with this possible over counting, the CHARMM [9] force field introduces a scaling parameter which allows one to reduce the strength of the 1-4 electrostatic and vdW interactions. At present, there is no physical justification to choose a particular value for such parameter. Nevertheless, in most large scale MD computations (e.g., see force field AMBER [10], and YAMBER [14], NAMD [15]) used for simulations of soft molecules (polymers and proteins) its value is arbitrary fixed. A priori it is not obvious that a universal value of such a parameter exists.

In this paper we present what is, to the best of our knowledge, the first attempt to gain an understanding of the 1-4 scaling influence on physical properties, as implemented in NAMD code that uses a CHARMM force field. Because of the complexity of proteins, we have chosen to focus first on much simpler physical systems: n-alkanes. We base our analysis looking at the ability of interaction models that differ by scaling parameter to reproduce two aspects of experimentally determined physical properties of monolayers of alkanes adsorbed on graphite: temperatures of melting, and the mechanism of the transition. The universality of scaling is tested by comparing the optimal scaling parameter determined for three alkanes of different length: hexane $C_6H_{14}$ ($l \sim$ 6.4 Å), dodecane $C_{12}H_{26}$ ($l \sim 13$ Å) and tetracosane $C_{24}H_{50}$ ($l \sim 28$ Å). Large scale (0.1 μs stabili-zation runs, 20-40 ns production runs) Molecular Dynamics (*NVT*, all atom representation) simulations were performed within the NAMD [15] package using a CHARMM2 [9] force field. The force field offers several options of exclusion and scaling policies for non-bonded interactions. In the present study, the exclusion/scaling options have been fixed as follows: any one of the 1-2, 1-3 and 1-4 intramolecular energy component can be scaled. In the case of so called *scaled 1-4* exclusion option, the 1-4 electrostatic interaction is modified by a factor SF ($0 < SF < 1$) and 1-4 van der Waals energy is rescaled down using a (slightly) modified 6-12 interaction parameters defined in the CHARMM2 force field. This latter option was applied in our calculation. At the same time the interactions between pairs 1-2 and 1-3 are totally excluded from non-bonded terms. We will show that any modification of electrostatic SF dramatically affects physical properties of alkanes and modifies the balance between the torsional and non-bonded energies [16].

Figure 1 shows the temperature dependence of the Lennerd-Jones (LJ) intermolecular energy of the adsorbed alkane monolayers calculated assuming either a large (SF = 0.1) or small scaling (SF = 0.9) of 1-4 electrostatics. Clearly, by scaling 1-4 electrostatics terms one shifts the position of the inflection point on energy curve that indicates the temperature of the melting transition. Additionally, the amplitude of this shift increases with the backbone length in a monotonic way and reaches a value of several tens of degrees for tetracosane, the longest alkane studied here. Furthermore, as will be discussed later, the most interesting result induced by scaling 1-4 interactions is the change of the microscopic *mechanism* of melting. First, it should be recalled and emphasized that melting of alkanes layers adsorbed on graphite is quite different from the traditional picture of melting. Alkane molecules are neither isotropic nor rigid; for these reasons, as it has been shown and discussed in recent papers [1-7], the melting of solid layers is a cooperative process in which the molecular layer looses both inter- and intra-molecular order. Such a transformation is driven by a progressive deformation of the internal structure of molecules.

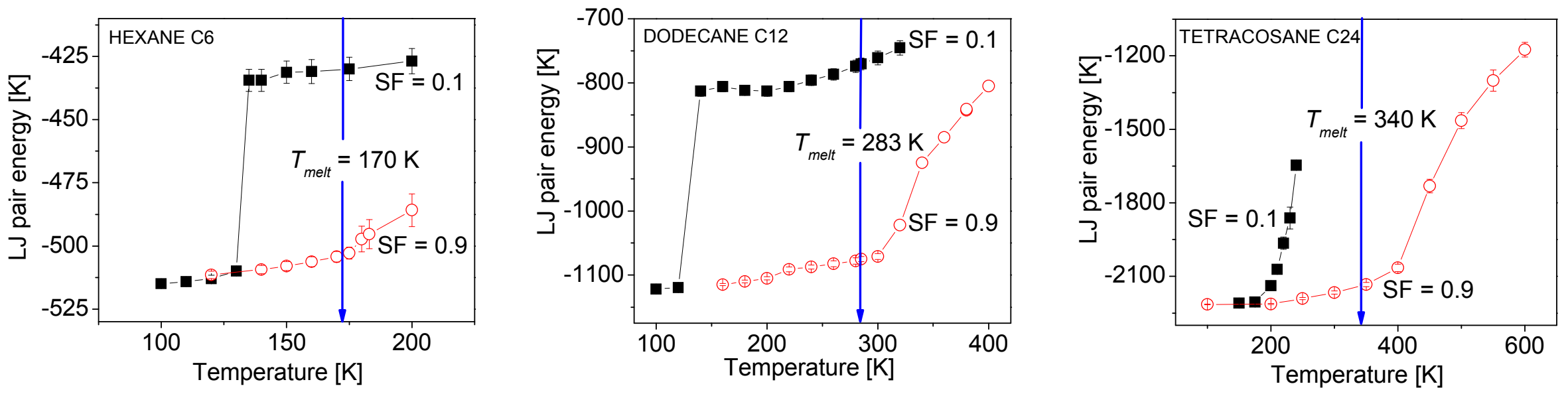

FIG. 1 (color online). LJ intermolecular energy as a function of temperature for two values of the scaling factor: SF = 0.1 and SF = 0.9. The inflection point indicates the melting transition in each model situation. The arrows indicate the experimental melting temperatures $T_{melt}$.

The rigidity of the molecular backbone may be characterized by its facility of gauche defect formation [6]. The apparent stiffness of molecules results from a competition between the internal forces and the external field, resulting from the interactions with neighboring molecules. Figure 2 shows the fraction of the gauche defects formed in a molecular backbone as a function of temperature, for different SF's. Several important conclusions can be inferred from this data: (i) The stiffness increases for higher SF's, meaning that the electrostatic term of the interaction stabilizes the linear configuration of molecules and prevents the backbone destabilization; (ii) The strength of intermolecular forces that defines topological aspects of melting increases with the molecule length. Therefore, both components tend to stabilize the solid state of the layers and in simulations the observed temperature of melting, $T_{melt}$, is higher than the experimental one for high SF's. These shifts increase with alkane length and for tetracosane reaches the unrealistic value of 200 K for SF = 1. On the other hand, if the 1-4 electrostatic interaction is almost totally suppressed (SF = 0.1) the simulated layers disorder easily via excessive chain deformation and melts at unrealistically low temperatures. Experimentally, the layers of shorter molecules melt with molecules much less deformed (stiffer), whereas in longer ones the deformation of the backbone appears to be the driving force of melting [1-7]. To reproduce this observation in computer simulations, the competition between the stiffness-defining components of energy should be correctly balanced. In this paper, we achieve this by scaling 1-4 non-bonded terms (in CHARMM2).

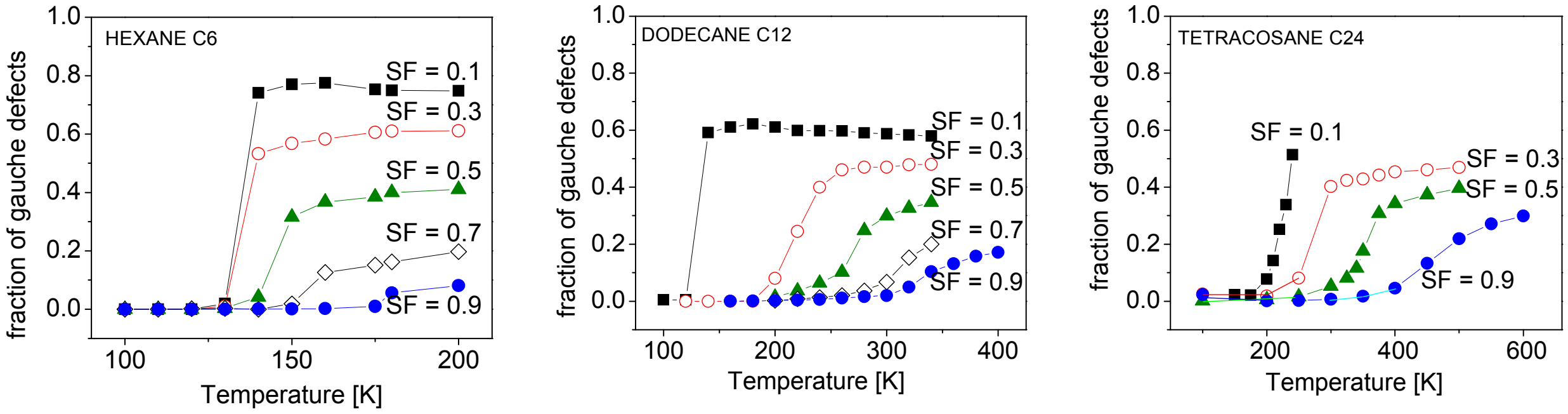

FIG. 2 (color online). Fraction of gauche defects formed within the molecular backbone at different SF's, as a function of temperature.

Logically, it could be possible to determine the correct SF directly from the calculations performed at the experimental $T_{melt}$. Figure 3 shows the intermolecular LJ energy calculated for each alkane at the experimental melting temperature (170 K, 283 K and 340 K for C6, C12 and C24, respectively) as a function of the SF. The optimal SF's determined as the inflection point at the curves in Fig. 3 are: SF ≅ 0.45 for tetracosane, SF ≅ 0.65 for dodecane and SF ≅ 0.8 for hexane. For optimal SF, all observables of the system (LJ energy, and also order parameters, see Fig. 4) show a clear singularity at $T_{melt}$. Figure 5 shows the variation of the optimal SF as a function of the number of carbon atoms in the alkane chain. This dependence can be tentatively approximated, within the error bars, by a monotonically decaying function which permits the estimation of the SF for other alkanes. Verification of our conclusions for other alkanes will be possible when experimental melting of other alkanes becomes known.

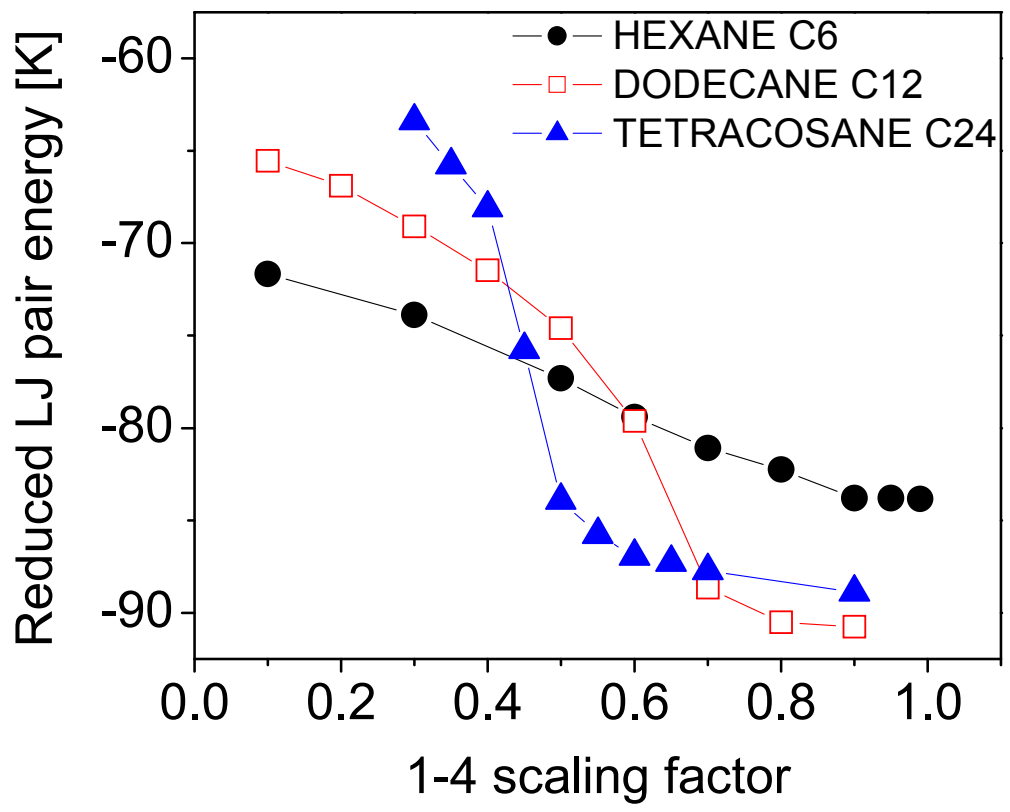

FIG. 3 (color online). LJ intermolecular energy as a function of 1-4 scaling factor. The calculations were done at the experimental temperatures of melting. For comparison, the values of LJ energy on the graph were normalized with respect to the number of carbons in the alkane chain.

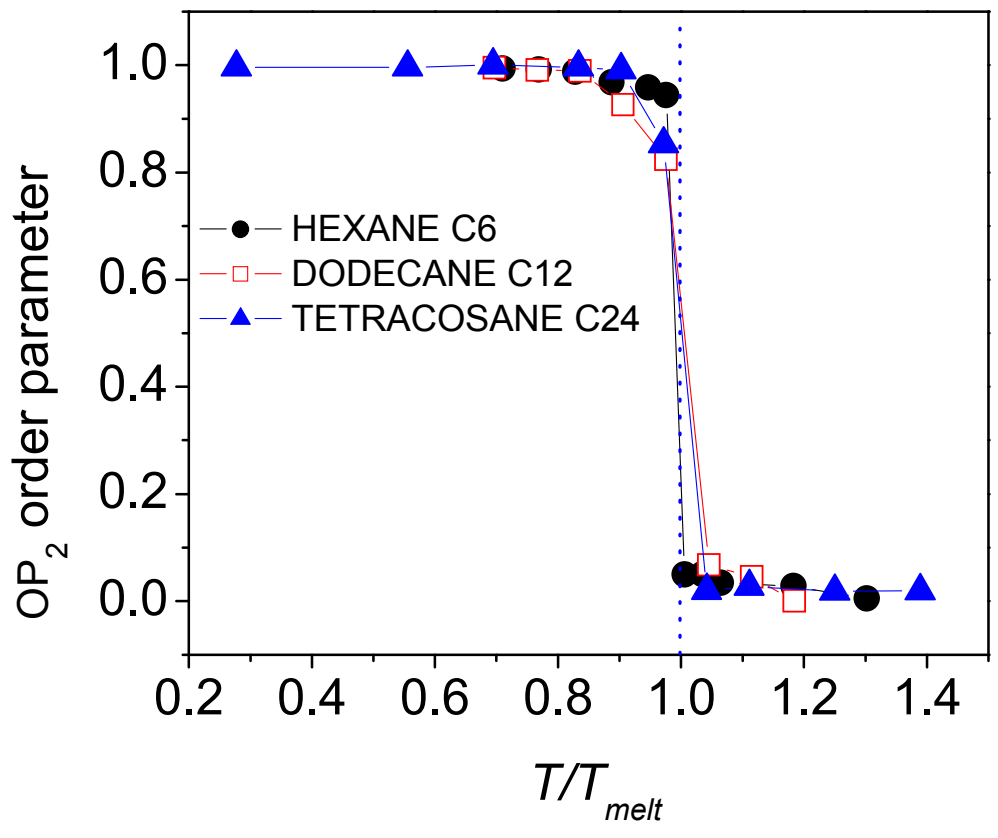

FIG. 4 (color online). $OP_2$ order parameter calculated for optimal scaling factors versus reduced temperature $T/T_{melt}$. $OP_2 = \sum_{i=1}^{N_m} \langle \cos 2\varphi_i \rangle / N_m$ , where the sum runs over all $N_m$ molecules in the simulation and $\phi_i \in [0,180°]$ is the angle that the smallest moment of inertia axis for molecule $i$ makes with the $x$-axis of the simulation box [3].

Figure 3 illustrates another interesting feature: each alkane shows different sensitivity to the scaling procedure. For C6 at $T_{melt}$ the LJ energy is not strongly dependent on the 1-4 SF. This is because in short alkanes ($n < 10$) melting is not induced by the internal deformation of molecules; therefore the conformational terms of intramolecular are less coupled to intermolecular forces and do not play significant role in the mechanism of melting. This conclusion is also supported by analysis of the average bond length (calculated from the end-to-end length, see Fig. 6). The change in the average bond length reflects the change of the molecular conformation at melting. Hexane molecules are stiff and remain unfolded up to the melting temperature, where the gauche defects start to form at the molecules ends [17]. The process of melting in the layer is thus driven by the reduction of molecules' footprint via rolling of the alkane plane to the orientation perpendicular with respect to the substrate and, subsequently, the molecules promotion to the second layer [3]. For long molecules, the deformation of molecular backbone through gauche defect formation (in-chain melting) assures the reduction of the molecules' footprint prior to intermolecular disordering. If a similar mechanism were responsible for melting in shorter alkanes, the melting temperature would be much lower (Figs. 1 and 2). Therefore for C6 the experimental temperature of melting is reproduced in the simulation only if the SF is relatively large. As a consequence, C6 molecules remain stiff and not deformed until the melting temperature. In the case of C12 and C24, much smaller SF's are required to correctly describe the smaller molecular stiffness.

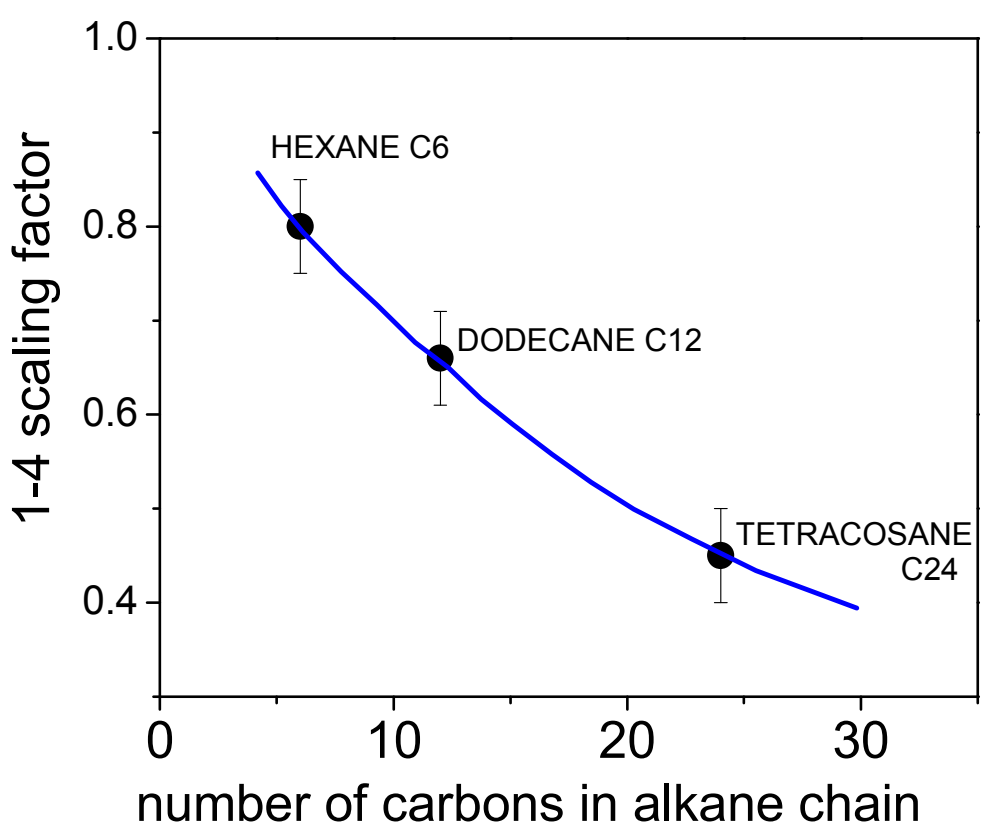

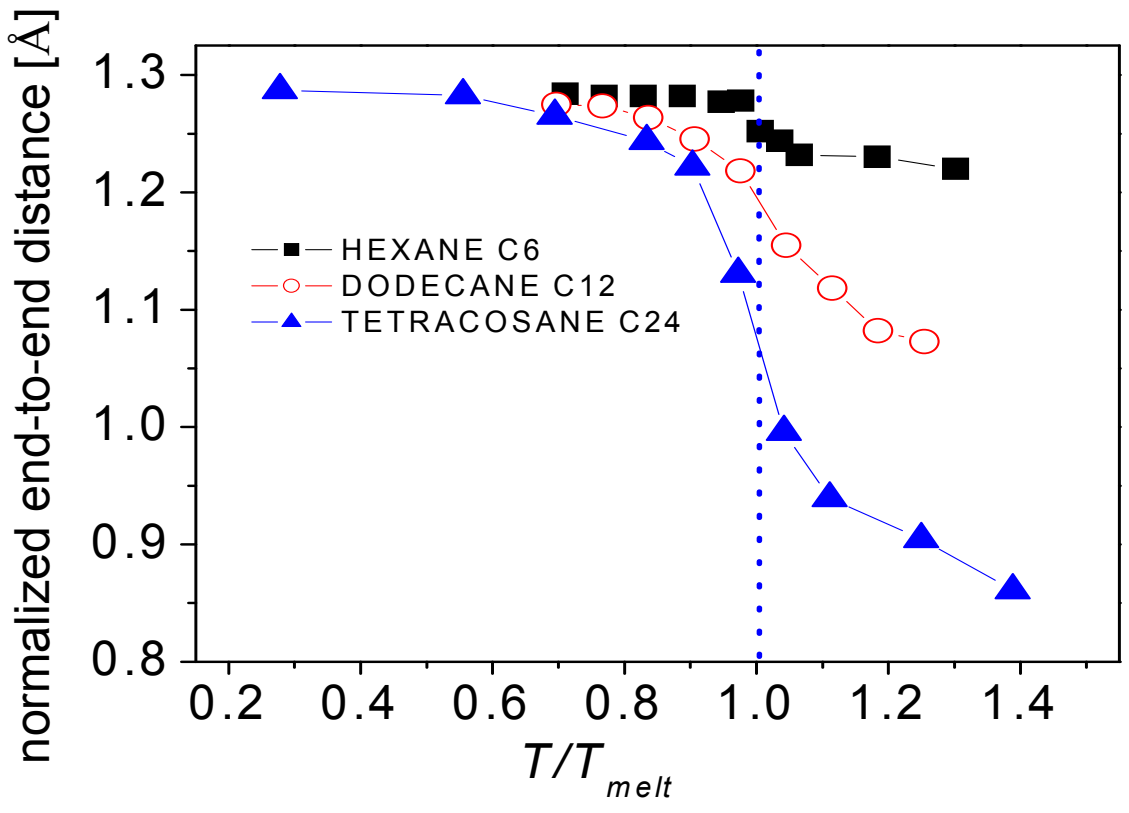

FIG. 5 (color online). Optimal scaling factor as a function of the number of carbons in the alkane backbone.

FIG. 6 (color online). Average end-to-end length of alkanes as a function of reduced temperature. Optimized scaling factors have been used in the calculations. The end-to-end distance was normalized with respect to the backbone number of C-C bonds.

Such observations are of extreme importance: there is a crossover between two limits of scaling 1-4 non-bonded interactions which modifies the melting mechanism. Depending on the alkane length, it is a continuous crossover between melting induced by the lattice instability (short alkanes, $n < 6$) to melting induced by instability of internal configurations of long molecules ($n \geq 12$).

From the point of view of computer modeling methodology, it is always desirable to have a set of *universal* force field parameters readily available that could be used for simulations of a broad range of materials. Unfortunately, for the SF of the 1-4 non-bonded interactions there is no theoretical formula which would allow estimating its value *a priori*. Very often a conventional value of 0.5 has been used in studies, even though there have been indications that this value may be not universal [10,14]. Here, we have explicitly shown that the scaling of the 1-4 non-bonded interactions is strongly system-dependent, even within a group of molecules that differ only in length. We have observed that, in the case of alkanes the scaling factor decreases when the number of carbons in the molecular backbone increases. Our results allow the estimation of the optimal value of SF for any linear alkane. Extrapolating our results to longer alkanes, we estimate the SF approaches zero for alkanes with more than 50 carbons in the backbone. Such inference is consistent with the fact that the CHARMM force field, developed for modeling of large biological molecules was parameterized without any scaling of the internal electrostatic energy [9,15].

The first important conclusion from this paper is the fact that the SF modifies profoundly physical properties of the system. The backbone torsional potential and 1-4 term scaling act collectively to determine the preferred conformational regions of the physical space and define the melting transition temperature. In consequence the mechanism of the melting is different when the scaling factor changes. When the SF is large (close to 1) the molecules are stiff and the melting of the adsorbed layer is induced by molecules promotion to the second layer. When SF decreases, the molecules are more flexible and the "footprint reduction" becomes the leading mechanism of the melting due to a higher probability of the gauche defect creation.

The second important conclusion of this work is the observation that electrostatic interactions may play a significant role for non-polar molecules and that the required value of the scaling factor is not universal. Any universality claim would lead to unrealistic mechanisms of the molecular deformation and eventually to melting mechanisms and temperatures very different from the experimental fact. Our work, however, raises the possibility that there could exist a universal function determining the SF value for families of molecules. The final verification of this dependence requires confrontation with more experimental data and, in particular, high quality experimental melting temperatures which are not presently available. A central unanswered, important question is how we should understand the physical basis of scaling of 1-4 interaction components in computer simulations. A simple explanation would relate it to the necessity of avoiding a double-counting of the interaction already included in the bond energies. It is interesting to note that the scaled energy is a relatively small part of the total system energy; however, it still plays an extremely important role in the systems' dynamics. Such an observation suggests that the usual treatment of electrostatic energy, limited only to the accounting for the static distribution of charges over the molecule may not, in fact, be correct. The first step to improve it would consist in an inclusion in the interaction model of the polarization energy and variable charges, depending on the conformation of the molecule. There are some reports indicating that both electrostatic and polarization energies may be of similar importance [17]. Such an effect can be important even in non-polar systems such as alkanes because of the large configurational fluctuations at melting which make instantaneous configurations of molecules highly asymmetric. Therefore application of charge distributions should be re-examined: the influence of polarization energy remains to be tested. This aspect is presently being studied and will be reported in the future.

**Acknowledgment**
The authors would like to thank H. Taub, F.Y. Hansen and R.D. Etters for enlightening discussions. Acknowledgment is made to the donors of The American Chemical Society Petroleum Research Fund (PRF43277-B5) for the support of this research. This material is based upon work supported in part by the Department of Energy under award number DE-FG02-07ER46411. Computational resources were provided by the University of Missouri Bioinformatics Consortium.